\documentclass[letterpaper, 10 pt, conference]{ieeeconf}  

\IEEEoverridecommandlockouts                              

\usepackage[utf8]{inputenc}
\usepackage{textcomp}
\usepackage[compatibility=false]{caption}
\usepackage{subcaption}
\usepackage{graphicx}
\usepackage{mathtools}
\usepackage{amssymb}
\usepackage{amsmath}
\usepackage{bm}
\usepackage[version=4]{mhchem}
\usepackage{siunitx}
\usepackage{longtable,tabularx}
\usepackage{algorithm}
\usepackage{algpseudocode}
\usepackage{url}
\usepackage{multirow}

\usepackage{amsthm}

\newtheoremstyle{sansindent}
  {3pt}
  {3pt}
  {\itshape}
  {}
  {\bfseries}
  {}
  { }
  {}

\theoremstyle{sansindent}
\newtheorem{prop}{Proposition}
\newtheorem{defn}{Definition}
\newtheorem{lem}{Lemma}

\newtheoremstyle{remark}
  {3pt}
  {3pt}
  {\normalfont}
  {}
  {\itshape}
  {.} 
  { }
  {}
  
\theoremstyle{remark}
\newtheorem{rem}{Remark}

\title{\LARGE \bf\resizebox{\textwidth}{!}
{Worst-case search in constrained uncertainty space for robust $\mathcal H_\infty$ synthesis}}

\author{Ervan Kassarian$^{1}$, Francesco Sanfedino$^{2}$, Daniel Alazard$^{2}$, Andrea Marrazza$^{1}$
\thanks{$^{1}$DYCSYT, 3 Avenue Didier Daurat, 31400 Toulouse, France.}%
\thanks{$^{2}$Fédération ENAC ISAE-SUPAERO ONERA, Université de Toulouse, 10 Avenue Marc Pélegrin, Toulouse, 31400, France}%
}

\begin{document}

\maketitle
\thispagestyle{empty}
\pagestyle{empty}

\begin{abstract}
Standard linear $\mathcal H_\infty$/$\mathcal H_2$ robust control and analysis tools operate on uncertain parameters assumed to vary independently within prescribed bounds.
This paper extends their capabilities in the presence of nonlinear constraints coupling these parameters and restricting the parametric space.
Based on the theory of upper-$C^1$ functions, it is shown that the sequential quadratic programming (SQP) algorithm can be slightly adapted to address the search for worst-case $\mathcal H_\infty$ norm, a nonsmooth constrained optimization problem, and the search for worst-case stability under some assumptions.
Specifically, we prove that for such upper-$C^1$ functions, any subgradient provides a descent direction and satisfies Karush-Kuhn-Tucker (KKT) conditions at a local minimum, and that any accumulation point generated by SQP is a KKT point.
This worst-case search then enables robust controller synthesis using a standard active configurations approach.
Through an application to the robust control of a satellite, the proposed approach is shown to provide a scalable framework for robustness analysis and robust controller synthesis.
\end{abstract}

\section{Introduction}

The $\mathcal H_\infty$ norm is a widely used metric to quantify the performance of control systems. 
In linear systems, minimization of the $\mathcal H_\infty$ norm was historically addressed with linear matrix inequalities (LMI). However, motivated by the limitations of LMIs for addressing large size problems and the practical need for structured controllers (e.g. fixed order), nonsmooth optimization was investigated, leading in particular to Matlab's function systune \cite{apkarian2006,Apkarian2006b} and to the open source toolbox HIFOO \cite{Burke2006a,Burke2006} . Thanks to their broad applicability to real-world systems, these techniques have found widespread use in industry across a wide range of applications, cf. e.g.~\cite{Frechard2012,Falcoz2015,Ganet2017,Serrano2021,Rahman2022}. They also allow to include $\mathcal H_2$ criteria \cite{apkarian2009,Arzelier2010}. 

However, practical problems include parametric uncertainties that may severely degrade the performance and the stability if not addressed during controller optimization.
Consequently, the robust structured $\mathcal H_\infty$ or $\mathcal H_2$ control problem can be formulated as the min-max constrained optimization problem:
\begin{equation} \label{eq:Hinf_multiobj}
 \begin{array}{lll}
    & \underset{K \in \mathcal K}{\text{minimize}} & \underset{\delta \in \mathcal D}{\text{max \;}}  || T_{z_1w_1}(s, K,\delta) ||  \\
    & \text{subject to } &  \underset{2 \leq i \leq n}{\max } \; \underset{\delta \in \mathcal D}{\text{max \;}}  || T_{z_iw_i}(s, K,\delta) || \leq 1
 \end{array}
\end{equation}
where $|| \cdot ||$ may refer to the $\mathcal H_2$ or $\mathcal H_\infty$ system norm, $T_{z_iw_i}(s, K,\delta)$ is a closed-loop transfer function from input $w_i$ to output~$z_i$, the structured controller $K$ belongs to a set $\mathcal K$, and $\delta$ is the vector of $k$ uncertain parameters in a parametric space~$\mathcal D$. 
Classically in the literature, $\mathcal D$ is the hypercube $[-1,1]^m$ after normalization of the $m$ uncertain parameters.
In this context, the nonsmooth optimization based approach of \cite{apkarian2006} was generalized to address the synthesis of a robust controller, first with $\mu$-synthesis~\cite{Apkarian2015} and then with an approach based on the construction of a set of active configurations~\cite{apkarian2015a}.
Specifically, for a fixed controller, \cite{apkarian2015a} solves the inner maximization subproblems associated with Problem~\eqref{eq:Hinf_multiobj} with local nonsmooth optimization, based on the computation of Clarke subdifferentials~\cite{Clarke90} of the $\mathcal H_\infty$ norm~\cite{Boyd91}. Worst-case search in stability is also addressed by maximizing the spectral abscissa.
Problem~\eqref{eq:Hinf_multiobj} is then tackled by combining this worst-case search with the non-smooth controller optimization of~\cite{apkarian2006}, as follows. This approach relies on the construction of an inner approximation $\mathcal D_a \subset \mathcal D$ composed of so-called active configurations.
At each iteration, the controller is tuned so as to be robust to all current active configurations. Then, a worst-case search is performed; if the worst-case configuration does not satisfy the optimization constraints, it is added to $\mathcal D_a$ and the algorithm goes back to controller tuning; otherwise, the algorithm can terminate.
It is shown that this approach is both more reliable and more efficient than gridding the parametric space, which quickly becomes intractable when the number of parameters increases and may miss critical configurations between grid points.
In practice, the optimization generally finishes with a small number of active configurations.

Since this approach relies on local searches, robustness is generally verified a posteriori using complementary techniques. For this purpose, Monte Carlo sampling provides statistical confidence with respect to the worst-case configuration found among a given number of samples~\cite{Tempo1996}. Motivated by the need to characterize the global worst case, $\mu$-analysis~\cite{doyle1982} was developed within the standard framework where $\mathcal D=[-1,1]^m$ and is generally based on branch-and-bound algorithms~\cite{roos2013systems}.
It provides guaranteed deterministic lower and upper bounds on the global optimum, although its scalability is limited by the NP-hardness of the problem~\cite{Toker98}.
It is worth noting that, even in the context of $\mu$-analysis, there is an interest in optimization-based techniques for computing lower bounds of the worst-case performance. In particular,~\cite{Casati2025} considers the $\mathcal H_2$ norm, which is differentiable with respect to the uncertain parameters, and reports that such approaches can significantly reduce computational cost compared with the more traditional LMI-based techniques, even for relatively small-scale systems.
However, the case of the $\mathcal H_\infty$ norm is more challenging due to its nonsmoothness.

The goal of this paper is to address robustness analysis problems where the admissible uncertainty domain is a subset of $[-1,1]^m$ defined by excluding certain regions.
Specifically, let $c_1(\delta), ..., c_n(\delta)$ denote a set of differentiable constraints defining the parametric space
\begin{equation} \label{eq:D}
   \mathcal D = \lbrace \delta \in [-1,1]^m \text{ s.t. } c_i(\delta) \leq 0 , \; 1 \leq i \leq n\rbrace \;.
\end{equation}
Since solving the worst-case search problems enables the robust control problem~\eqref{eq:Hinf_multiobj} to be addressed using the approach of~\cite{apkarian2015a}, this paper focuses on worst-case search for the $\mathcal H_\infty$ norm:
\begin{equation} \label{eq:wc_perfo}
\underset{\delta \in \mathcal D}{\max }  || T_{zw}(s, K,\delta) ||_\infty \;.
\end{equation}
Worst-case stability analysis is also investigated, as it is implicitly required to ensure that the norms remain finite. Following~\cite{apkarian2015a,Apkarian2017wc}, it is formulated as
\begin{equation} \label{eq:wc_stab}
\underset{\delta \in \mathcal D}{\max } \alpha(A(K,\delta)),
\end{equation}
where $A(K,\delta)$ denotes the state matrix of the closed-loop system and $\alpha(\cdot)$ denotes the spectral abscissa.
These two problems are addressed through local optimization. The rationale for this choice is twofold. First, we seek computationally efficient techniques that can be integrated into robust control design, alternating between multi-model controller synthesis and worst-case search as in~\cite{apkarian2015a}, in order to solve Problem~\eqref{eq:Hinf_multiobj} within reasonable computational time. Second, as a validation tool, we aim to complement conventional global exploration techniques, for example by enhancing Monte Carlo simulations through the addition of a local search to improve the detection of rare worst-case configurations, or by extending existing local-search-based approaches used in $\mu$-analysis for lower-bound computation.

Problems~\eqref{eq:wc_perfo} and~\eqref{eq:wc_stab} are widely recognized as nontrivial due to their nonsmoothness. Nevertheless,~\cite{Noll2013,apkarian2015a} relate the $\mathcal H_\infty$ norm, viewed as a function of a structured controller or uncertainty, to the theory of lower- and upper-$C^1$ functions~\cite{Spingarn81}, a class of nonsmooth functions with several favorable regularity properties~\cite{Rockafellar1982}. Under the assumption of simple eigenvalues, the spectral abscissa can also be characterized as such a function~\cite{apkarian2015a,Bompart2007}.
Multiple references suggest that the minimization of upper-$C^1$ functions, even when not differentiable
everywhere, exhibits similar behavior to smooth functions.
In particular, in \cite{apkarian2015a}, the authors show the convergence of their bundle method even when the cutting plane step is reduced to a standard linesearch (as in smooth optimization) when applied to such functions. This work follows~\cite{Dao2015} which deals with the optimization of lower- and upper-$C^1$ functions. Similar considerations are also thoroughly discussed by the same authors in~\cite{Apkarian2017wc} in the context of robust control.
In~\cite{Oliveira2023}, the Frank-Wolfe algorithm, originally designed for smooth functions, is also proven to converge to Clarke-stationary points when applied to upper-$C^1$ functions (in the context of unconstrained optimization).

In this paper, we demonstrate new properties of the class of upper-$C^1$ functions for constrained optimization, and in particular that any subgradient verifies Karush-Kuhn-Tucker (KKT) conditions when the constraint functions $c_i$ are differentiable. We also propose a slight adaptation of the Sequential Quadratic Programming (SQP)~\cite{Nocedal2006}, a smooth constrained optimization algorithm, to the upper-$C^1$ case, and demonstrate the convergence to KKT points. We deduce that this algorithm can be applied to the nonsmooth problem~\eqref{eq:wc_perfo}, and to \eqref{eq:wc_stab} under an additional assumption.
Following the framework introduced in~\cite{apkarian2015a}, the proposed worst-case search method is then integrated with established multi-model controller optimization approaches~\cite{apkarian2006} to enable robust controller synthesis.
The proposed methodology is illustrated through the control of a realistic mechanical system, which typically involves constraints of the form~\eqref{eq:D}. The results demonstrate that the proposed local optimization strategy complements Monte Carlo sampling to identify more consistently rare worst-case configurations. Furthermore, the proposed approach enables efficient robust controller optimization and exhibits good scalability to plants with a large number of states and uncertain parameters, owing to its gradient-based formulation.

 \section*{Notations}

\noindent $T_{zw}(s)$ denotes a transfer function between an input $w$ and an output $z$. 
$\alpha(\cdot)$ denotes the spectral abscissa:
\begin{equation*}
    \alpha(M) \vcentcolon = \max \lbrace \textrm{Re} (\lambda): \lambda \text{ is an eigenvalue of } M \rbrace \;.
\end{equation*}
Given  $f : \mathbb R^n \to \mathbb{R}$ a locally Lipschitz function, the Clarke subdifferential of $f$ at point $x$ is denoted $\partial f(x)$.
When it exists, the directional derivative of $f$ at $x$ in direction $d$ is defined as 
\[
f'(x,d) = \lim_{t \to 0^+} \frac{f(x+td) - f(x)}{t} \;.
\]

\section{Constrained optimization of upper-$C^1$ functions}

\label{sec:optim}

\subsection{Properties of upper-$C^1$ functions}

In this section, we introduce the class of upper-$C^1$ functions and demonstrate favorable properties for their minimization under constraints. As we shall see in Section~\ref{sec:2} in the context of robust $\mathcal H_\infty$ control, the worst-case search in performance~\eqref{eq:wc_perfo}, and in stability~\eqref{eq:wc_stab} under an additional assumption, can be addressed with such functions.

\vspace{2mm}

\begin{defn} \textbf{\emph{-- Upper- and lower-$C^1$ functions \cite{Spingarn81}}} \label{def:upperC1}

 A locally Lipschitz function $f : \mathbb R^n \to \mathbb{R}$ is \emph{upper-$C^1$} at $x \in \mathbb R^n$ if there exists a compact set $S$, a neighborhood $U$ of $x$, and a function $F : U \times S \to \mathbb{R}$ such that $F$ and $\nabla_x F$ are jointly continuous in $x$ and $s$, and 
\[
  f(x) = \min_{s \in S} F(x',s) \quad \text{for all } x' \in U.
\]
We denote $I(x)$ the subset of $S$ that defines the active functions $F(\cdot,s)$ at $x$, i.e.:
\[
I(x) = \arg \min_{s \in S} \; F(x,s) = \lbrace s \in S, f(x) = F(x,s) \rbrace \;.
\]
$f$ is \emph{lower-$C^1$} if $-f$ is upper-$C^1$, i.e. $f$ can be written as a maximum, similarly to the expression above.
\end{defn}

In this whole section, we consider an upper-$C^1$ function $f: \mathbb R^n \rightarrow \mathbb R$.
We recall that the directional derivative always exists for an upper-$C^1$ function \cite{Rockafellar1982}.
Let us first consider the unconstrained problem:
\begin{equation} \label{eq:unconstrained_problem}
\underset{x \in \mathbb R^n}{\text{minimize}} \; f(x)
\end{equation}

\vspace{2mm}

\begin{prop} 
\textbf{\emph{-- Descent directions and optimality of upper-$C^1$ functions -- Unconstrained case}}\label{prop:upperC1unconstrained}

\noindent Let a locally Lipschitz function $f$ be upper-$C^1$ at $x$.

    \noindent \emph{a) Descent direction. }
    Given  $g_x \in \partial f(x)$ any subgradient such that $||g_{x}||\neq0$,  $-g_{x}$ is a descent direction for $f(x)$, and in particular it verifies $f'(x, -g_{x})\leq - || g_{x} ||^2 < 0$.

    \noindent \emph{b) Optimality. }
    If $x$ locally minimizes $f(x)$, then  $||g_{x}||=0$ for any subgradient $g_x \in \partial f(x)$.

\end{prop}

\begin{proof}
a) \cite[Proposition 2 and equation (6)]{Oliveira2023} shows that the directional derivative of an upper-$C^1$ function $f$ is:
\begin{equation} \label{eq:directional_deriv}
    f'(x, d) = \min_{g \in \partial f(x)} g^T d = \min_{u \in I(x)}  \nabla_x F(x, u)^T d
\end{equation}
From Eq.~\eqref{eq:directional_deriv}, $f'(x, -g_{x}) = \min_{g \in \partial f(x)} (-g^Tg_{x}) \leq -g_{x}^Tg_{x} = - || g_{x} ||^2 \leq 0$ and the last inequality is strict when $|| g_{x} || \neq 0$.

\noindent b)
If $x$ minimizes $f$, then $f'(x,-g_{x})\geq 0$.
But, by a), $f'(x,-g_{x}) \leq - || g_{x} ||^2$. Thus, $|| g_{x}||=0$. 
\end{proof}

Proposition~\ref{prop:upperC1unconstrained}.a) highlights a favorable property of upper-$C^1$ functions, namely their similarity with $C^1$ functions, for which any gradient provides a descent direction. Proposition~\ref{prop:upperC1unconstrained}.b) further shows that an optimization algorithm can rely on the same first-order optimality conditions as in the smooth case when selecting any subgradient. This should be contrasted with the general nonsmooth case, where subgradients do not generally define a descent direction, and where the optimality condition is $0 \in \partial f(x)$, a weaker condition commonly referred to as Clarke stationarity.

We shall now demonstrate similar results for the constrained problem:
\begin{equation}  \label{eq:upperC1problem}
 \begin{array}{lll}
    & \underset{x \in \mathbb R^n}{\text{minimize}} & f(x)  \\
    & \text{subject to } &  c_i(x) \leq 0 \; , \;\; 1 \leq i \leq n 
 \end{array}
\end{equation}
where the $c_i$ are differentiable, and $f$ is upper-$C^1$.
We do not consider equality constraints, as they can be written equivalently as two inequality constraints.
The Karush-Kuhn-Tucker (KKT) conditions of Problem~\eqref{eq:upperC1problem} read: there exists $g_x \in \partial f(x)$ and $u=(u_i)_{1 \leq i \leq n}$ such that:
\begin{equation} \label{eq:kktoptim}
\begin{array}{l}
    g_{x} + \sum_{i=1}^n u_i \nabla c_i(x) = 0 \\
    \forall i, \; c_i(x) \leq 0 \\
    \forall i, \; u_i \geq 0 \\
    \forall i, \; u_i c_i(x) = 0
\end{array}
\end{equation}
We also define the quadratic problem $Q(x,H,g_x)$:
\begin{equation} \label{eq:qp}
     \begin{array}{lll}
    & \underset{p \in \mathbb R^n}{\text{minimize}} & g_x^Tp + \frac{1}{2} p^THp  \\
    & \text{subject to } &  c_i(x) + \nabla c_i(x)^Tp \leq 0 \; , \;\; 1 \leq i \leq n 
 \end{array}
\end{equation}
where $H$ is any symmetric positive definite matrix, and $g_x \in \partial f(x)$. The quadratic problem $Q(x,H,g_x)$ was originally introduced in~\cite{Han1997}, with the gradient $\nabla f(x)$ instead of the subgradient $g_x$. It can be seen as a local approximation of the Problem~\eqref{eq:upperC1problem} where the matrix $H$ may contain information about the curvature of both objective and constraint functions, usually serving as an approximation of the Lagrangian. The KKT conditions of $Q(x,H,g_x)$ are: there exists $u=(u_i)_{1 \leq i \leq n}$ such that:
\[ \begin{array}{l}
g_{x} + \sum_{i=1}^n u_i \nabla c_i(x) + \frac{1}{2} (H+H^T)p = 0\\
\forall i, \; c_i(x) + \nabla c_i(x)^Tp \leq 0 \\
\forall i, \; u_i \geq 0 \\
\forall i, \; u_i \geq 0 , \; u_i (c_i(x) + \nabla c_i(x)^Tp) = 0 
\end{array}
\]

Finally, for $r \geq 0$, we introduce the \emph{merit function} $\theta_r$ as in \cite{Han1997}:
\begin{equation} \label{eq:meritfcn}
    \theta_r(x) = f(x) + r \sum_{i=1}^n c_i(x)_+ 
\end{equation}
where $c_i(x)_+ = \max(0, c_i(x))$.

\vspace{2mm}

\begin{prop}\textbf{\emph{-- Descent directions and optimality of upper-$C^1$ functions -- Constrained case}}\label{prop:upperC1constrained}

\noindent Let a locally Lipschitz function $f$ be upper-$C^1$ at $x$. Let the $c_i$ be continuously differentiable.

    \noindent \emph{a) Descent direction. }
    Let $(p,u)$ be a KKT pair of $Q(x,H,g_{x})$ with $g_{x} \in \partial f(x)$ any subgradient.
    If $||p||\neq0$ and $|u|_\infty \leq r$ (vector $\infty$-norm), then $p$ is a descent direction for $\theta_r(x)$, i.e. $\theta'_r(x,p) < 0$.

    \noindent \emph{b) Optimality. } If $x$ locally minimizes $f(x)$ subject to $n$ constraints $c_i(x) \leq 0$ (Problem \eqref{eq:upperC1problem}), and if a constraint qualification holds at $x$ for the active constraints, then any subgradient $g_{x} \in \partial f(x)$ verifies the following KKT conditions: there exists $(\mu_i)_{1\leq i \leq n}$ such that:
    \begin{equation*}
\begin{array}{l}
    g_{x} + \sum_{i=1}^n \mu_i \nabla c_i(x) = 0 \\
    \forall i, \; \mu_i \geq 0 , \; \mu_i c_i(x) = 0, \; c_i(x) \leq 0
\end{array}
\end{equation*}
\end{prop}

\begin{proof}
    a) Cf. Appendix. The proof is a mere adaptation of \cite[Theorem 3.1]{Han1997} to the upper-$C^1$ case.

    \noindent b) 
    Let us note $\mathcal C = \lbrace x , c_i(x) \leq 0 \; \forall i \rbrace$. Since $x$ is a local minimizer of $f(x)$ in $\mathcal C$, there exists a ball $B(x,r_1)$ of radius $r_1>0$ and of center $x$ such that $f(x') \geq f(x)$ for all $x' \in B(x,r_1) \cap \mathcal C$. Recall that $U$ is the neighborhood of $x$ of Definition~\ref{def:upperC1}; as $B(x,r_1)$ and $U$ are both non-empty neighborhoods of $x$, we can define $r_2>0$ such that $B(x,r_2) \subset B(x,r_1) \cap U$. Let us consider any of the functions $F(\cdot , s)$. For all $x' \in B(x,r_2) \cap \mathcal C$, we have $F(x',s) \geq f(x') \geq f(x)$, where the first inequality follows from $x' \in U$, and the second from $x' \in B(x,r_1) \cap \mathcal C$. Assuming additionally that $F(\cdot , s)$ is active, i.e. $s \in I(x)$, then $F(x,s)=f(x)$ and we conclude that $F(x',s) \geq F(x, s)$ for all $x' \in B(x,r_2) \cap \mathcal C$.

This shows that $x$ locally minimizes all active functions $F(\cdot, s)$ under the constraints $c_i(x) \leq 0$. Furthermore, by assumption, the active constraints satisfy a constraint qualification at $x$.
Therefore, the functions $F(\cdot, s)$ satisfy the following KKT conditions: for all $s \in I(x)$, there exists $( \lambda_i(s))_{1\leq i \leq n}$ such that:
\begin{equation*}
\begin{array}{l}
    \nabla_x F(x,s) + \sum_{i=1}^n \lambda_i(s) \nabla c_i(x) = 0 \\
    \forall i, \; \lambda_i(s) \geq 0 , \; \lambda_i(s) c_i(x) = 0, \; c_i(x) \leq 0
\end{array}
\end{equation*}
As shown in~\cite{Clarke90,Oliveira2023}, the subdifferential of an upper-$C^1$ function $f$ at $x$ is the convex hull formed by the gradients of the active functions $F(\cdot, s)$:
\[
\partial f(x) = \operatorname{co} \left\{ \nabla_x F(x, s) : s \in I(x) \right\}
\]
Thus, $g_{x} \in \partial f(x)$ can be written as:
\[
g_{x} = \sum_{j=1}^p a_{s_j} \nabla_x F(x, s_j) \;, \;\; s_j \in I(x) \; \forall j
\]
where the $a_{s_j} \geq 0$ are such that $\sum_{j=1}^p a_{s_j}=1$ (and $p$ is at most the number of parameters plus one). Then, we have:
\begin{align*}
g_{x} = - \sum_{j=1}^p a_{s_j} \sum_{i=1}^n \lambda_i(s_j) \nabla c_i(x) = - \sum_{i=1}^n \left( \sum_{j=1}^p a_{s_j} \lambda_i(s_j) \right) \nabla c_i(x) =  - \sum_{i=1}^n \mu_i \nabla c_i(x)
\end{align*}
where $\mu_i := \sum_{j=1}^p a_{s_j} \lambda_i(s_j)$. For all $i$: $\mu_i \geq 0$ (because all $a_{s_j}$ and $\lambda_i(s_j)$ are $\geq 0$), and $\mu_i c_i(x) = 0$ (because if $c_i(x) \neq 0$, then all $\lambda_i(s_j)$ are 0). This proves Proposition~\ref{prop:upperC1constrained}.b).
\end{proof}

\begin{rem}
Proposition~\ref{prop:upperC1constrained}.a) is a generalization of \cite[Theorem 3.1]{Han1997}, originally for differentiable functions, to upper-$C^1$ functions. It will be used in next section to prove convergence of an optimization algorithm. Proposition~\ref{prop:upperC1constrained}.b) can be seen as a specific case of the usual KKT conditions for upper-$C^1$ functions. It allows stationarity to be tested with any subgradient, in contrast to the general nonsmooth case, where only the existence of one subgradient satisfying KKT conditions is required.
\end{rem}

\subsection{SQP algorithm for upper-$C^1$ functions}

We now introduce the Sequential Quadratic Programming (SQP) algorithm adapted for upper-$C^1$ functions to solve Problem \eqref{eq:upperC1problem}. It is described in Algorithm~\ref{sqp}, and is identical to that of the original reference~\cite{Han1997} for differentiable functions, except that the gradient $\nabla f(x)$ is replaced by any arbitrary subgradient $g_x \in \partial f(x)$ in the quadratic problem $Q(x,H,g_x)$ (Eq.~\eqref{eq:qp}). We do not detail how the quadratic problem in Step~1 or the step size $\lambda_k$ in Step~2 may be solved, or how $H_k$ is updated in Step~3, as it does not affect the convergence proof (as long as the conditions described in Algorithm~\ref{sqp} are satisfied).

\begin{algorithm}
\caption{SQP algorithm \cite{Han1997} adapted for upper-$C^1$ functions}\label{sqp}
\begin{algorithmic}
\State \textbf{Initialization}: $x_0 \in \mathbb R^n$, $H_0 \in \mathbb R^{n \times n}$ positive definite, $r > 0$, $\lambda_{max} > 0$.

\State \textbf{Step 1}: Pick any $g_{x_k}\in \partial f(x_k)$. Find a KKT point $p_k$ of the quadratic problem $Q(x_k, H_k, g_{x_k})$.

\State \textbf{Step 2}: Set:
\[x_{k+1} = x_k + \lambda_k p_k\]
and find a $\lambda_k \in [0, \lambda_{max}]$ satisfying: 
\[  \theta_r(x_{k+1}) \leq \min_{0 \leq \lambda \leq \lambda_{max}} \theta_r(x_k + \lambda p_k) + \epsilon_k\]
where $\theta_r$ is defined in Eq.~\eqref{eq:meritfcn}, and ($\epsilon_k$) is a sequence of nonnegative numbers satisfying:
\[ \sum_{i=0}^\infty \epsilon_i < \infty\]
\State \textbf{Step 3}: Update $H_{k+1}$ by some scheme that keeps it definite positive.
\end{algorithmic}
\end{algorithm}

To prove the convergence of Algorithm~\ref{sqp}, the following lemmas are necessary.

\vspace{2mm}

\begin{lem} \textbf{\emph{-- Supermonotonicity of the subdifferential of upper-$C^1$ functions}} \label{lemma1}
Let $f$ be upper-$C^1$ at $x$. For every $\epsilon > 0$, there exists $\delta > 0$ such that for all $x_0,x'_0 \in \mathcal B(x, \delta)$:
    \[
    (g_{x'_0} - g_{x_0})^T(x'_0-x_0) \leq \epsilon || x'_0 - x_0 ||
    \]
    where $g_{x_0} \in \partial f(x_0)$ and $g_{x'_0} \in \partial f(x'_0)$.
\end{lem}
Lemma~\ref{lemma1} comes from the submonotonicity of the subdifferential of lower-$C^1$ function \cite[Theorem 3.9]{Spingarn81} and is also noted for example in~\cite[Lemma 6]{Apkarian2016bundle}.

\vspace{2mm}

\begin{lem} \textbf{\emph{-- Perturbation of a quadratic problem}} \label{lemma2}
    Let us consider the two quadratic problems:
    \begin{equation*}
    x_0 = \arg \min \lbrace\; Q(x)=\frac{1}{2} x^T K x - x^T k \;\; 
    \text{subject to } \; Gx \leq g \rbrace
    \end{equation*}
    \begin{equation*}
    x_0' = \arg \min \lbrace\; Q'(x)=\frac{1}{2} x^T K' x - x^T k' \;\; 
    \text{subject to } \; G'x \leq g' \rbrace
    \end{equation*}
    where $K,K'$ are symmetric, positive definite matrices.
    Let us partition $G$ into two matrices $A,B$, and $g$ into two vectors $a,b$ such that $B x \leq b$ is always satisfied as an equality while $Ax \leq a$ can be satisfied with strict inequality. Let us define similarly $B'$ in the partition of $G'$ and suppose that $\mathrm{rank}(B) = \mathrm{rank}(B')$.
    Set 
    $$\epsilon = \max \left( || K - K'||, ||G-G'||, ||g-g'||, \frac{(k'-k)^T(x'_0-x_0)}{||x'_0-x_0||} \right)\;.$$
    Then there exists $\epsilon_0>0$ and $c>0$ such that:
    \[
    ||x_0'-x_0|| \leq c \epsilon
    \]
    whenever $\epsilon < \epsilon_0$.
\end{lem}
\begin{proof}
    Cf. Appendix. The proof is a mere adaptation of~\cite{Daniel1973}.  
\end{proof}

Finally, Proposition~\ref{th:convergence} extends the convergence result of SQP established in~\cite[Theorem 3.2]{Han1997} for differentiable objective functions to the case of upper-$C^1$ objective functions, when providing any subgradient where a gradient would be expected. Besides the upper-$C^1$ condition of $f$, the same assumptions as in~\cite[Theorem 3.2]{Han1997} are considered.

\vspace{2mm}

\begin{prop} \textbf{\emph{-- Convergence of SQP for upper-$C^1$ functions}} \label{th:convergence}

    Let $f$ be upper-$C^1$ and the $c_i$ be continuously differentiable. Assume that the following conditions are satisfied:

    (i) there exists $\alpha, \beta > 0$ such that, for each $k$ and for any $x \in \mathbb R^n$:
    \[  \alpha x^Tx \leq x^T H_k x \leq \beta x^Tx\]

    (ii) For each $k$, there exists a KKT point of $Q(x_k, H_k, g_{x_k})$ (Eq.~\eqref{eq:qp}) with a Lagrange multiplier vector $u_k$ such that $||u_k||_\infty \leq r$ (vector $\infty$-norm).

    Then, any sequence $(x_k)$ generated by Algorithm~\ref{sqp} either terminates at a KKT point of the constrained optimization problem~\eqref{eq:upperC1problem}, or any accumulation point $\bar x$ with
    \[  S^0(\bar x) = \lbrace p: c(\bar x) + \nabla c(\bar x)^Tp < 0 \rbrace \neq \emptyset \]
    is a KKT point of Problem~\eqref{eq:upperC1problem}.
\end{prop}

\begin{proof}
    The complete proof is provided in the Appendix and adapts the proof of~\cite[Theorem 3.2]{Han1997} to the case where $f$ is not differentiable but only upper-$C^1$. Here, we simply summarize the main adaptations:

    \noindent 1)
    When $f$ is differentiable, it is always possible to choose $x_{k+1}$ as described in Algorithm~\ref{sqp} thanks to \cite[Theorem 3.1]{Han1997}.
    This argument is maintained in the upper-$C^1$ case thanks to Proposition~\ref{prop:upperC1constrained}.a), which extends \cite[Theorem 3.1]{Han1997} from differentiable to upper-$C^1$ functions.

    \noindent 2)     
    When $f$ is differentiable, given $\bar x$ an accumulation point of $(x_k)$, the continuity of the gradient of $f$ ensures that $\nabla f(x_k)$ converges to $\nabla f(\bar x)$. In the upper-$C^1$ case, a similar property is maintained as follows. Since $f$ is locally Lipschitz, its subgradients $(g_{x_k})$ are uniformly bounded for $k$ sufficiently large, hence there exists a subsequence that converges to a vector $\bar g$. Without loss of generality (passing to the subsequence if necessary) we note
    \[  g_{x_k} \rightarrow \bar g \;,
    \]
    and the upper semi-continuity of the Clarke subdifferential \cite[Proposition 1.5]{Clarke1998} yields $\bar g \in \partial f(\bar x)$.

    \noindent 3)
    When $f$ is differentiable, given the convergence of the (sub)sequences ($x_k$), ($H_k$), ($g_{x_k}$) and ($\nabla c_i(x_k)$), the existence of an accumulation point of the solutions $p_k$ of the quadratic problems $Q(x_k,H_k,g_{x_k})$ follows from the continuity of $\nabla f(x)$ and from \cite{Daniel1973}. This argument is maintained in the upper-$C^1$ case thanks to Lemmas~\ref{lemma1} and~\ref{lemma2}, where $x_{k+1}$ and $x_k$ play the roles of $x_0$ and $x'_0$, and $g_{x_0}$ and $g_{x'_0}$ play the roles of $k$ and $k'$.   
\end{proof} 

Essentially, this shows that standard SQP implementations can be applied to upper-$C^1$ functions by providing an arbitrary subgradient wherever a gradient would normally be required. Indeed, Proposition~\ref{prop:upperC1constrained}.b) shows that, at a local minimum, the KKT conditions can be verified using any subgradient, as in the smooth case where the gradient is uniquely defined. Moreover, Proposition~\ref{th:convergence} establishes that SQP iterations converge to KKT points when an arbitrary subgradient is used. This result must be understood in the general nonsmooth sense: there exists at least one subgradient satisfying the KKT conditions, but these conditions need not be satisfied by all possible subgradients.

\begin{rem}
    The purpose of advocating for an adaptation of a smooth optimization technique is not to claim superiority over specialized nonsmooth constrained optimization methods, such as those developed for lower- or upper-$C^1$ functions in~\cite{Dao2016bundle,Dao2015}. Rather, we emphasize that readily available optimization algorithms, implemented for instance in standard Matlab or Python toolboxes, can be successfully applied.
    Nevertheless, when the matrix $H_k$ is updated with the Broyden-Fletcher-Goldfarb-Shanno (BFGS), the proposed algorithm may benefit from superlinear convergence rate if, after a finite number of iterations, the upper-$C^1$ function $f$ becomes differentiable and satisfies the assumptions of~\cite{Powell78a}.
    In the general case, however, even an upper-$C^2$ property might not be sufficient alone to guarantee superlinear convergence, since the active indices defining the max representation of $f$ may change infinitely often in a neighborhood of the solution.
\end{rem}

\section{Worst-case search and robust control in constrained space}

\label{sec:2}

\subsection{Worst-case search}

This section focuses on worst-case search for a fixed controller $K$, which is therefore omitted from the notation for simplicity.
We consider a generic transfer function written classically as the Linear Fractional Transformation (LFT) \cite[Chapter 10]{Zhou1996} $T_{zw}(s, \delta) = \mathcal F_u(M(s),\Delta)$, as depicted in Fig.~\ref{fig:lft}, where $\Delta = \textrm{diag} (\delta_i I_{n_i})$ contains real uncertain parameters $\delta_i$, assumed without loss of generality to belong to $[-1,1]$, with $\delta_i=0$ corresponding to the nominal configuration. The integer $n_i$ denotes the number of repetitions of the uncertainty $\delta_i$.
\begin{center}
\includegraphics[width=.2\linewidth]{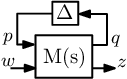}
\captionof{figure}{LFT representation of $T_{zw}(s, \delta)$}
\label{fig:lft}
\end{center}
The state-space representation of $M(s)$ is written as:
\begin{equation*}
\begin{array}{l}
\left[
\begin{array}{c}
    \dot x \\ \hline q \\ z
\end{array}\right]
    =
\left[
\begin{array}{c|cc}
    A_1 &  B_1 & B_2 \\ \hline
    C_1 & D_{11} & D_{12} \\
    C_2 & D_{21} & D_{22}
\end{array}\right]
\left[
\begin{array}{c}
    x \\ \hline p \\ w
\end{array}\right]
\end{array}
\end{equation*}

The performance problem \eqref{eq:wc_perfo} is rewritten as the constrained optimization problem:
\begin{equation} \label{eq:hinf_problem}
\begin{array}{lll}
    & \underset{\delta \in \mathbb R^n}{\text{minimize}} &  h_\infty(\delta) : = - || T_{zw}(s, \delta) ||_\infty  \\
    & \text{subject to }&  c_i(\delta) \leq 0  \\
\end{array}
\end{equation}
where the constraints $c_i (\delta)$ are assumed to be differentiable, and include the usual bounds $-1 \leq \delta_i \leq 1$ but may also contain other (possibly nonlinear) constraints, which is the main focus of this paper.


\vspace{2mm}

\begin{prop} \textbf{\emph{-- Regularity of $h_\infty(\delta)$ \cite{Boyd91,apkarian2006,apkarian2015a}}} \label{prop:hinfregu}

\noindent a) $h_\infty(\delta)$ is locally Lipschitz and Clarke subdifferentiable (in the sense of \cite{Clarke90}) everywhere in 
    \[\mathcal D_s = \lbrace \delta \in \mathcal D,  T_{zw}(s, \delta) \text{ is stable} \rbrace \;.
    \]
    \noindent b) $h_\infty(\delta)$ is upper-$C^1$ everywhere in $\mathcal D_s$.
    
    \noindent c) $h_\infty(\delta)$ is differentiable in $\mathcal D_s$ when only one frequency is active and of multiplicity one; in this case, its gradient is given in Eq.~\eqref{eq:grad_h_}.
\end{prop}

Proposition~\ref{prop:hinfregu}.a) is proved in~\cite[Proposition 1]{apkarian2015a} and follows the work of \cite{Boyd91} and \cite{apkarian2006}. The general subdifferential expression is provided in these references, but will not be used in this paper.
b) is proved in~\cite[Lemma 9]{Noll2013}  or in~\cite[Proposition 2]{apkarian2015a}.
c) immediately follows from a), and the expression is also provided in~\cite[Remark 5]{apkarian2015a}: using the same notations, let us define $T_{qw}$ and $T_{zp}$ as:
\begin{equation*}
\left[
    \begin{array}{c|c}
        * & T_{qw} (s, \delta) \\ \hline
        T_{zp} (s, \delta) & T_{zw}(s, \delta)
    \end{array}
    \right] = 
    \left[
    \begin{array}{c|c}
        0 & I \\ \hline
        I & \Delta
    \end{array}
    \right] \star M(s)
\end{equation*}
with $\star$ denoting the redheffer star product. The transfer functions $T_{zw}$, $T_{qw}$ and $T_{zp}$ correspond to the transfer between the inputs $p,w$ and outputs $q,z$ of the matrix $M(s)$ as in Fig.~\ref{fig:lft}, but after closing the loop with the block $\Delta$.
If $\omega_0$ is the single active frequency, such that the multiplicity of $\bar \sigma(T_{zw}(j\omega_0, \delta))$ is 1, $h_\infty$ is differentiable and its derivatives read:
\begin{equation} \label{eq:grad_h_}
    \frac{\partial h_\infty}{\partial \delta_i} = -\mathrm{Tr} \left( \mathrm{Re}(T_{qw}(j\omega_0, \delta) p_{\omega_0} q_{\omega_0}^H T_{zp}(j\omega_0, \delta))^T \Delta_i \right)
\end{equation}
where $p_{\omega_0}$ and $q_{\omega_0}$ are the normalized right and left singular vectors of $T_{zw}(j\omega_0, \delta)$ associated with $\bar \sigma(T_{zw}(j\omega_0, \delta))$, and $\Delta_i := \frac{\partial \Delta}{\partial \delta_i}$ \cite{apkarian2015a}.

From the results of Section \ref{sec:optim} and Proposition~\ref{prop:hinfregu}.b), we deduce that we can use SQP (Algorithm~\ref{sqp}) to solve Problem~\eqref{eq:hinf_problem}.
In particular, let us define the direction $d_\delta$ such that its components are defined as in Eq.~\eqref{eq:grad_h_} (even when $h_\infty$ is not differentiable), i.e. the i-th component of $d_\delta$ reads:
\begin{equation*}
    d_{\delta,i} = -\mathrm{Tr} \left( \mathrm{Re}(T_{qw}(j\omega_0, \delta) p_{\omega_0} q_{\omega_0}^H T_{zp}(j\omega_0, \delta))^T \Delta_i \right)
\end{equation*}
where $\omega_0$ is an active frequency -- but not necessarily the only one, and the multiplicity of $\bar \sigma(T_{zw}(j\omega_0, \delta))$ is not necessarily 1~-- and $p_{\omega_0}$ and $q_{\omega_0}$ are the corresponding normalized right and left singular vectors of $T_{zw}(j\omega_0, \delta)$. 
Then, $d_\delta$ is a subgradient of $h_\infty$ (this is straightforward from the expression of the subdifferential given in~\cite[Section IV-1]{apkarian2015a}); by Propositions~\ref{prop:upperC1constrained} and~\ref{th:convergence}, it can legitimately be used in SQP as if it were the gradient of a differentiable function.

 \begin{rem}
 In this context, Proposition~\ref{prop:upperC1constrained}.b) (optimality condition for upper-$C^1$ functions) formalizes the intuitive observation that a local maximum of the $\mathcal H_\infty$ norm is also a local maximum of all active peaks under the considered constraints. This situation is likely to occur in practice, as multiple active frequencies often arise after controller optimization. Conversely, Proposition~\ref{prop:upperC1unconstrained}.a) (descent direction for upper-$C^1$ functions in the unconstrained case) indicates that increasing the gain of at least one active peak, by moving along a subgradient, necessarily increases the $\mathcal H_\infty$ norm. Proposition~\ref{prop:upperC1constrained}.a) extends this result to the constrained case, where the merit function is considered instead.
\end{rem}

\begin{rem}
In contrast, minimizing the lower-$C^1$ function $-h_\infty(\delta)$ (e.g. finding a controller that minimizes the $\mathcal H_\infty$ norm), with or without constraints, would be a more complex, genuinely nonsmooth problem. In particular, a local minimum may occur at a point where many subgradients are nonzero, as the point may not correspond to a minimum of any of the active peaks individually. Moreover, a descent direction cannot be obtained by selecting an arbitrary subgradient, since decreasing one active peak may be compensated by the increase of another, leading to an increase of the $\mathcal H_\infty$ norm. For this problem, nonsmooth techniques in the literature include multidirectionnal search \cite{apkarian2006MDS}, gradient sampling \cite{Burke2005}, or Clarke subdifferential-based bundle methods \cite{apkarian2006,Bompart2007a}.
\end{rem}


As in~\cite{apkarian2015a}, worst-case stability is studied by maximizing the spectral abscissa of the closed-loop matrix $A(\delta)$. Problem~\eqref{eq:wc_stab} can also be written as the following constrained optimization problem:
\begin{equation} \label{eq:stability_problem}
\begin{array}{lll}
    & \underset{\delta \in \mathbb R^n}{\text{minimize}} & a(\delta) : = - \alpha(A(\delta)) \\
    & \text{subject to }& c_i(\delta) \leq 0 
\end{array}
\end{equation}
where the constraints $c_i (\delta)$ are differentiable, as in Problem~\eqref{eq:hinf_problem}.
The case of the spectral abscissa is less favorable than the $\mathcal H_\infty$ norm. We recall that simple eigenvalues (i.e. those of multiplicity one) are differentiable \cite{Overton1988}. ``Active eigenvalues" designates all eigenvalues whose real part is equal to the spectral abscissa.

\vspace{2mm}

\begin{prop} \textbf{\emph{-- Regularity of $a(\delta)$ \cite{apkarian2015a,Bompart2007}}} \label{prop:clarke_a}

\noindent a) If all active eigenvalues of $A(\delta)$ are semi-simple (i.e. their algebraic and geometric multiplicity are equal), then $a(\delta)$ is locally Lipschitz and Clarke subdifferentiable at $\delta$.

\noindent b) If all active eigenvalues of $A(\delta)$ are simple at $\delta$, then $a(\delta)$ is upper-$C^1$ at $\delta$.

\noindent c) If $A(\delta)$ has only one active eigenvalue and it is simple, then $a$ is differentiable at $\delta$. The derivative of $a(\delta)$ with respect to $\delta_i$ reads:
\begin{equation} \label{eq:grad_a_}
    \frac{\partial a}{\partial \delta_i} = -\mathrm{Tr} ( \Delta_i^T \mathrm{Re}((I-D_{11} \Delta)^{-1} C_1 v_i  u_i^H B_1 (I-\Delta D_{11})^{-1})^T )
\end{equation}
where $v_i$ and $u_i$ are right and left eigenvectors associated with the active eigenvalue, normalized with $u_i^H v_i=1$. It also holds if there are two active eigenvalues and they are complex conjugate eigenvalues \cite{Bompart2007}.
\end{prop}

Proposition~\ref{prop:clarke_a}.a) was established in~\cite{apkarian2015a} following the earlier work of \cite{Bompart2007}.  These references build on \cite{Burke94} which studied the directional derivatives of the spectral abscissa. b) is also shown in~\cite{apkarian2015a}, and simply follows from the fact that the spectral abscissa is the maximum of the real parts of the eigenvalues, which are differentiable when the eigenvalues are simple \cite{Overton1988}. For c), the expression \eqref{eq:grad_a_} is straightforward from the expression of the Clarke subdifferential given in~\cite{apkarian2015a}.

\begin{rem}
Although the assumption of semi-simplicity is required for the definition of the Clarke subdifferential, the numerical experiments reported in~\cite{apkarian2015a} show that their bundle approach, based on Clarke subdifferentials, performs well in practice even without explicitly verifying this assumption.
\end{rem}

\begin{rem}
As for the $\mathcal H_\infty$ norm, minimizing the spectral abscissa is a more complex problem; in the literature, this is addressed with bundle methods \cite{Oustry2000,Burke2002,Noll2005,Bompart2007}, or gradient sampling \cite{Burke2005}.
\end{rem}

In this paper, as for $h_\infty(\delta)$, we rely on the upper-$C^1$ property of $a(\delta)$ of Proposition~\ref{prop:clarke_a}.b) and on the results of Section~\ref{sec:optim}, when active eigenvalues are simple. However, the case of non-simple active eigenvalues is admittedly a limitation of of the proposed approach for the worst-case search in stability.

\subsection{Robust control in constrained parametric space}

\label{sec:3}

Algorithm \ref{algo} follows the approach proposed in~\cite{apkarian2015a} that relies on iteratively constructing a subset $\mathcal D_a \subset \mathcal D$ of so-called active configurations to solve Problem~\eqref{eq:Hinf_multiobj}. Here, the worst-case search developed in the previous sections is integrated to address constrained parametric spaces of the form~\eqref{eq:D}. Other minor practical differences with respect to~\cite{apkarian2015a} are detailed in Section~\ref{sec:practical}. 

The set of active configurations is initialized with the nominal configuration.
Step~1 optimizes the controller for all current active configurations. This step is not addressed in this paper; the most common available algorithm to address it is~\cite{apkarian2006}. 
Step~2 addresses the worst-case search in stability as described in Section~\ref{sec:2}, based on SQP (Algorithm~\ref{sqp}). If a Karush-Kuhn-Tucker (KKT) point $\delta^*$ of the optimization problem is found that leads to a spectral abscissa greater than a tolerance $\alpha_{max}$, it is added to $\mathcal D_a$. It is also possible to identify multiple KKT points during Step~2, as discussed in Section~\ref{sec:practical}.
Step~3 tackles the worst-case search in performance using:
\begin{equation} \label{eq:multiobj2}
     p(K, \delta) =  \max\; \left\lbrace \frac{1}{(1+\epsilon_1)\beta}  || T_{z_1w_1}(s, K,\delta) || \;, \right. \left. \frac{1}{(1+\epsilon_2)}\; \underset{2 \leq i \leq n}{\text{max \;}}   || T_{z_iw_i}(s, K,\delta) || \right\rbrace 
\end{equation}
where, for completeness, $||\cdot ||$ may denote either the $\mathcal H_\infty$ or the $\mathcal H_2$ norm (even though only the former requires the developments of this paper), and where $\beta$ is the performance obtained in Step 1:
\begin{equation*}
    \beta =  \underset{\delta \in \mathcal D_a}{\text{max \;}}  || T_{z_1w_1}(s, K^*,\delta) || \;.
\end{equation*}
The stopping criterion in Step~3 is that $p(K, \delta)$ must be less than 1, i.e. we consider that the algorithm has converged when the worst-case search degrades the soft requirement by less than $\epsilon_1 \%$, and the hard requirements are less than $1+\epsilon_2$.

\begin{algorithm}
\caption{Robust control optimization with constrained uncertainties}\label{algo}
\begin{algorithmic}
\vspace{1mm}
\State \textbf{Initialization}: Initialize the set of active configurations as $\mathcal D_a = \lbrace 0 \rbrace $.
\vspace{1mm}
\State \textbf{Step 1}: Multi-model synthesis.
\State Compute optimal controller over active configurations:
\begin{equation*}
 \begin{array}{rrl}
    & K^*=\arg \; \min_{K \in \mathcal K} & \underset{\delta \in \mathcal D_a}{\text{max \;}}  || T_{z_1w_1}(s, K,\delta) ||  \\
    & \text{subject to } &  \underset{2 \leq i \leq n}{\max } \; \underset{\delta \in \mathcal D_a}{\text{max \;}}  || T_{z_iw_i}(s, K,\delta) || \leq 1
 \end{array}
\end{equation*}
\vspace{1mm}
\State \textbf{Step 2}: Worst-case search in stability.
\State Identify KKT point(s) $\delta^* = \arg \; \max_{\delta \in \mathcal D} \; \alpha(A(K^*,\delta))$.
\State If $\alpha(A(K^*,\delta^*)) > \alpha_{max}$, add $\delta^*$ to $\mathcal D_a$ and go back to step 1.
Otherwise continue to step 3.
\vspace{1mm}
\State \textbf{Step 3}: Worst-case search in performance.
\State Identify KKT point(s) $\delta^* = \arg \; \max_{\delta \in \mathcal D} \; p(K^*,\delta)$.
\State If $p(K^*,\delta^*) > 1 $, add $\delta^*$ to $\mathcal D_a$ and go back to step 1.
Otherwise, terminate.
\end{algorithmic}
\end{algorithm}

\subsection{Practical aspects}

\label{sec:practical}



\noindent \textit{Monte Carlo sampling and SQP}

From a practical perspective, rather than initializing SQP from the nominal configuration, which may converge towards local optima, a global exploration of the parameter space can be performed beforehand. In this paper, Monte Carlo sampling is considered for this purpose. Since the objective is to address worst-case search without probabilistic consideration, it is natural to sample the parameters from uniform distributions.

In this paper, two implementations are investigated experimentally. The first consists in running SQP from a given number of randomly sampled initial points. Denoting $f(\delta)$ a function of the random variable $\delta$ (in our case, either $a(\delta)$ or $h_\infty(\delta)$), the second consists in randomly sampling a number $N$ of samples $\{\delta_1,\ldots,\delta_N\}$ satisfying:
\begin{equation*}
    N \geq \frac{\textrm{ln}(\gamma)}{\textrm{ln}(1-\epsilon)}
\end{equation*}
where $\gamma$ and $\epsilon$ serve as confidence and accuracy levels to guarantee the probability \cite{Tempo1996}:
\begin{equation} \label{eq:MonteCarlo}
    \mathbb P_{\delta_1,... \delta_N} ( \mathbb P_\delta ( f(\delta) < \hat f_N) \leq \epsilon ) \geq 1-\gamma
\end{equation}
where $\hat f_N = \min_{1 \leq i \leq N} f(\delta_i)$. 
Note that the two probability levels are required because $\hat f_N$ is a random variable induced by the Monte Carlo samples $\{\delta_1,\ldots,\delta_N\}$. Consequently, the outer probability is evaluated with respect to the sampling process generating $\{\delta_1,\ldots,\delta_N\}$, while the inner probability is computed with respect to the underlying distribution of $\delta$.
This bound is independent of the number of uncertain parameters.
Then, a local search with SQP is performed starting from the worst-case sample.
Alternatively, it is also possible to draw $M$ batches of $N/M$ samples and to initialize a SQP with the worst-case sample of each batch; the stochastic guarantee is the same as in Eq.~\eqref{eq:MonteCarlo} (tied to the worst-case sample among the total of $N$ samples) but it allows to perform multiple local searches from different points of the parametric space to increase coverage. If those SQP runs can be parallelized, this procedure comes at no extra cost.

\vspace{2mm}

\noindent \textit{Adding one vs multiple KKT points to active set}

Algorithm~\ref{algo}, along with the discussion above, allows one to add multiple KKT points to the active configurations after a worst-case search is performed.
This paragraph elaborates on how to select which KKT points to add when multiple candidates are available, e.g. after running several SQP in parallel.

Consider two KKT points $\delta_1^*$ and $\delta_2^*$. If their distance $|| \delta_1^* - \delta_2^* ||$ is below a predefined threshold, adding both points to the active configuration set would increase the size of the optimization problem without any meaningful benefit.
This threshold can typically be chosen as a fraction of the diagonal of the hypercube $2\sqrt{n}$, where $n$ is the number of uncertain parameters.
However, this approach may not be robust to the presence of uninfluential parameters, which may have very different values in $\delta_1^*$ and $\delta_2^*$ and thus contribute to the distance, without actually justifying to include both configurations to the active set.
This limitation can be circumvented by weighting the distance with sensitivity indices $w_i$, computed for each parameter. For this purpose, global sensitivity analysis methods \cite{Saltelli2004} generally only require a few hundreds of samples, and thus come at limited extra cost. Then, the distance is defined as $|| W (\delta_1^* - \delta_2^*) ||_2$ with $W=\text{diag(}w_i$). The threshold may now be chosen as a fraction of $2 ||w||_2$ the diagonal of the rescaled hyperrectangle.

Although this sensitivity-based criterion is heuristic, it does not affect the validity of the algorithm. Its only impact is on computational efficiency: an excessively large threshold may introduce unnecessary active configurations, whereas an excessively small threshold reduces the method to adding configurations one at a time, as in the original approach of~\cite{apkarian2015a}.
Conversely, a well-chosen threshold can reduce the overall computational cost by decreasing the number of calls to the costly Step~1.

\section{Illustration: robust control of a satellite}

\label{sec:4}

The benchmark used in this study is designed to be realistic in terms of complexity (number of states, uncertain parameters, and decision variables), challenging regarding robustness, and includes a nonlinear constraint on the uncertain parameters that is typical of mechanical systems. The tests were run on a 12th Gen Intel(R) Core(TM) i7-12700H 2.30 GHz, with 14 cores and 16GB of RAM, on Matlab R2025a. Computation times are provided for reference, but they may vary depending on hardware, software configuration, implementation details, optimization parameters, etc.

\subsection{Benchmark model}

We consider a spacecraft, composed of one rigid body connected to two large flexible solar panels.
Uncertainties are considered on various mechanical properties; in total, there are 43 uncertain parameters $\delta_i$, and the $\Delta$ block of the LFT has size 204. 
The spacecraft model $G(s, \delta)$ is generated with the Satellite Dynamics Toolbox of \cite{Alazard2020,Sanfedino2023sdt}.
Denoting by $M_P(\delta)$ the mass matrix at the attachment point $P$ of the solar panel, and by $L_P(\delta)$ the matrix of modal participation factors obtained from a finite element modal analysis, the robustness analysis should exclude any configuration $\delta$ such that the residual mass matrix $M_r(\delta)$ defined in Eq.~\eqref{eq:Mr} is not positive definite (indeed, such configurations do not represent a physical system). Hence the nonlinear constraint:
\begin{equation} \label{eq:Mr}
    M_r(\delta) := M_P(\delta) - L_p^T (\delta) L_p(\delta) \succeq 0
\end{equation}
which can also be written under a scalar form:
\begin{equation} \label{eq:constraint}
    c(\delta) : = \alpha(-M_r(\delta)) \leq 0
\end{equation}
and is differentiable as long as deficiency occurs along a single principal direction in this application.
The LFT model includes all configurations for $-1 \leq \delta_i \leq 1$ even though only $\sim22\%$ of them respect the nonlinear constraint \eqref{eq:constraint}.

\begin{figure}[h!]
\centering
\includegraphics[width=.4\linewidth]{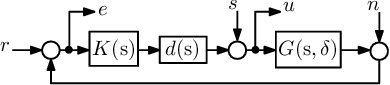}
\captionof{figure}{Benchmark: closed-loop system}
\label{fig:closedloop}
\end{figure}

The benchmark focuses on the attitude control of the satellite described above to follow a reference $r$. The control input $u$ of the plant $G(s, \delta)$ is the torque applied to the satellite, and the output vector contains the angle and rate measurements polluted by some noise $n$.
The closed loop is represented in Fig.~\ref{fig:closedloop}, where $d(s)$ represents actuator dynamics.
We denote $\mathcal K$ the set of controllers $K(s)$ of the desired structure: one 4-th order controller per spacecraft axis, resulting in 24 decision parameters per controller with Matlab's default setting (72 decision parameters in total).
The closed-loop plant has 50 states.

The control problem consists of placing a control bandwidth of 0.1 rad/s while ensuring a modulus margin of 0.5 and minimizing the variance of the actuator command in response to measurement noise. It reads:
\begin{equation} \label{eq:control_problem}
\begin{array}{lll}
    & \underset{K \in \mathcal K}{\text{minimize}} & \underset{\delta \in \mathcal D}{\max } \; || T_{un} (s, K, \delta) ||_2^2 \\
    & \text{subject to } &  \underset{\delta \in \mathcal D}{\max } \; || \frac{1}{2} T_{us} (s, K, \delta) ||_\infty < 1 \\
    & & \underset{\delta \in \mathcal D}{\max } \; || W(s) T_{er} (s, K, \delta) ||_\infty < 1  \\
    & & \underset{\delta \in \mathcal D}{\max } \; \alpha(A(K,\delta)) < -10^{-7}
\end{array}
\end{equation}
where $\mathcal D$ is defined in \eqref{eq:D} with the nonlinear constraint \eqref{eq:constraint}, the factor $\frac{1}{2}$ in the 2nd requirement ensures the desired modulus margin, and $W(s) = \frac{10 s + 1}{2(10 s + 0.01)} I_3$ is a weighting filter to impose the desired bandwidth of 0.1 rad/s. The last requirement of Problem~\eqref{eq:control_problem} ensures closed-loop stability with some tolerance, where $A(K,\delta)$ is the matrix $A$ of the state-space representation of the closed-loop system.

Solving Problem~\eqref{eq:control_problem} with Algorithm~\ref{algo} involves worst-case search problems of the form given in~\eqref{eq:hinf_problem} ($\mathcal H_\infty$ performance) and~\eqref{eq:stability_problem} (stability), subject to the nonlinear constraint~\eqref{eq:constraint}. Note that the worst-case search of the $\mathcal H_2$ norm is also included: it is handled similarly, even though it does not even require the developments presented in this paper, since the $\mathcal H_2$ norm is differentiable wherever it is well defined (see, e.g.,~\cite{Rautert1997,apkarian2009,Arzelier2010}).

Such a control problem is typical of satellites applications, cf. e.g.~\cite{Falcoz2015}, and cannot be tackled by current robust control tools because of the constraint~\eqref{eq:constraint}.

\subsection{Worst-case search}

\label{sec:4_wc_search}

In this section, we illustrate the worst-case search for the spectral abscissa, as presented in Section \ref{sec:2}.
We selected a controller $K_1$ obtained at an initial iteration of the robust controller tuning presented in Section~\ref{sec:4_control}, and analyzed the robustness of the closed-loop system.
We search for $\max_{\delta \in \mathcal D} \; \alpha(A(K_1,\delta))$.

\vspace{2mm}

\noindent \textit{SQP with random initial points}

\begin{figure}[ht] 
  \begin{subfigure}[b]{0.5\linewidth}
    \centering
    \includegraphics[width=.5\linewidth]{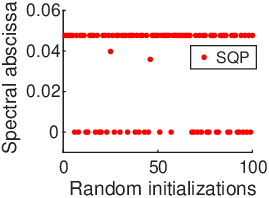} 
    \caption{SQP alone} 
    \label{fig:wc_abscissa} 
  \end{subfigure}
  \begin{subfigure}[b]{0.5\linewidth}
    \centering
    \includegraphics[width=.5\linewidth]{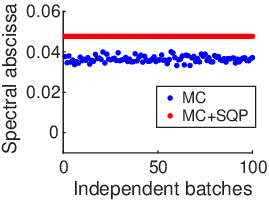} 
    \caption{MC and MC+SQP} 
    \label{fig:wc_abscissa_mc} 
  \end{subfigure}
  \label{fig:testcase1} 
  \caption{Worst-case search results (spectral abscissa)}
\end{figure}

We ran the SQP algorithm implemented in Matlab's function fmincon\footnote{As termination criteria, we set $10^{-6}$  as optimality tolerance (KKT stationarity condition), and $10^{-8}$ as step tolerance (step size between two iterations).} from 100 random initial points. The resulting worst-case spectral abscissa of each run are in Fig.~\ref{fig:wc_abscissa}, and show the presence of multiple local optima.
The computational time per run was 4.7s on average, with a maximum of 51s for the longest case.

\vspace{2mm}

\noindent \textit{Monte-Carlo + SQP}

We sampled 100 batches of $N=460$ samples ($\epsilon=\gamma=0.01$ in Eq.~\eqref{eq:MonteCarlo}), and selected the worst sample of each batch to initialize an SQP run. Fig.~\ref{fig:wc_abscissa_mc} presents the worst sampled spectral abscissa for each batch (in blue) and the worst case after the subsequent SQP (in red). Overall, MC+SQP performs consistently better than MC alone. Moreover, even the worst MC sample across all 100 batches (46000 samples in total) is below the worst case consistently detected by SQP.
Each group took 0.57s on average (with parallel evaluation of the samples); the SQP runtimes were between 3s and 9s. Thus, the Monte-Carlo step comes at little to no extra cost (if parallel computing is available), and can be an interesting alternative to the random initializations of SQP.

\subsection{Robust control}

\label{sec:4_control}

Finally, the robust control Problem \eqref{eq:control_problem} was solved using Algorithm~\ref{algo} with tolerances $\epsilon_1=5\%$ and $\epsilon_2=1\%$ (cf. Eq.~\eqref{eq:multiobj2}). 
Step~1 was executed with Matlab's systune.
The worst-case searches performed in Steps~2 and~3 follow the procedure described in Section~\ref{sec:4_wc_search}. For the Monte Carlo approach, the parameters $\epsilon=\gamma=0.01$ are used in Eq.~\eqref{eq:MonteCarlo}, resulting in $N=460$ samples distributed over $M=10$ batches. After termination, Steps~2 and~3 are repeated once with a larger sampling budget: $\epsilon=\gamma=0.001$, yielding $N=6910$ samples. If robustness is invalidated during this additional verification step, the procedure is restarted from Step~1. At each iteration, it was allowed to add multiple KKT points to the set of active configurations, relying on the activity scores of~\cite{Constantine2017} to compute the weighted distances as discussed in Section~\ref{sec:practical}.

\begin{figure}[ht!]
\centering
\subfloat[$T_{un} (s, K^*, \delta)$]{\label{a}\includegraphics[width=.3\linewidth]{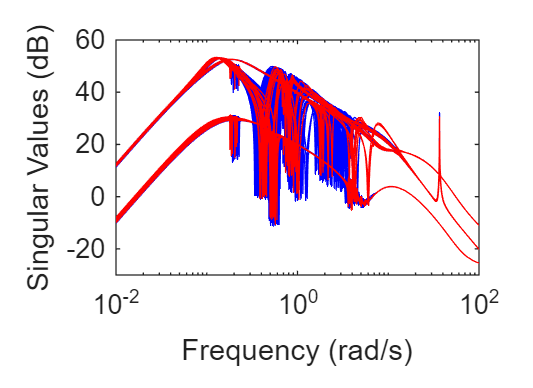}}\hfill
\subfloat[$T_{us} (s, K^*, \delta)$]{\label{b}\includegraphics[width=.3\linewidth]{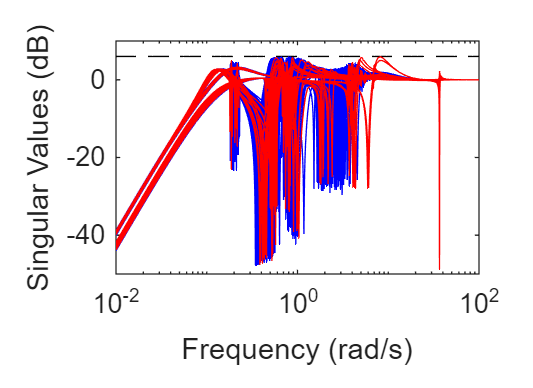}}\hfill 
\subfloat[$T_{er} (s, K^*, \delta)$]{\label{c}\includegraphics[width=.3\linewidth]{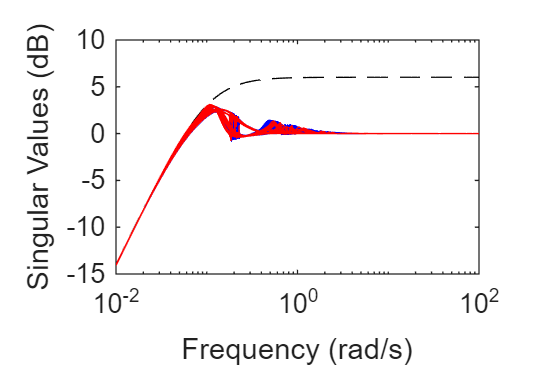}}
\caption{Singular values of the transfer functions in Problem \eqref{eq:control_problem} after controller optimization, with active configurations (red), 100 random samples (blue) and requirement (dashed lines)}
\label{fig:optim_results}
\end{figure}

For evaluation purposes, since the algorithm involves randomness through Monte Carlo sampling, the experiment was repeated 10 times, with the resulting metrics reported in Table~\ref{table:results}. All runs were initialized with the same nominal controller and converged to a robust controller. However, a significant variance is observed, due to the fact that active configurations are not necessarily discovered in the same order across different tests, because of the Monte Carlo step. The most computationally demanding phase is the controller optimization (Step~1).
Fig.~\ref{fig:optim_results} presents the transfer functions of Problem~\eqref{eq:control_problem} after optimization of the controller.

\begin{table}[ht!]
    \centering
    \normalsize
   \caption{Controller optimization results}
    \label{table:results}
   \begin{tabular}{| l || c | c | c |}
     \hline 
     Metrics & Min. & Average & Max. \\ \hline \hline
     Nb of iterations & 6 & 8.4 & 12   \\ \hline
     Nb of active configurations & 12  & 16.3 & 24   \\
     \hline
     Total runtime (s)   & 1312 &  2257  & 3954 \\ \hline
     ... including Step 1* runtime  & 66\% & 74\% & 85\% \\ \hline
   \end{tabular}
   
   * Controller optimization (systune), as opposed to the worst-case searches (MC+SQP, Steps 2 and 3)
 \end{table}


\section{Conclusion}

This paper focused on the synthesis of robust controllers and worst-case search in the presence of nonlinear constraints restricting the uncertainty space, a specificity that is missing from the current state of the art.

Relying on the theory of upper-$C^1$ functions, we established several properties favorable to the constrained optimization of such nonsmooth functions, namely that (i) any subgradient defines a valid descent direction, (ii) local optima satisfy the Karush–Kuhn–Tucker (KKT) conditions for all subgradients, and (iii) Sequential Quadratic Programming (SQP) converges to KKT points when applied to such functions. In control systems, this theoretical foundation legitimizes the use of standard, readily available optimization tools to perform worst-case search in performance, and in stability under some assumptions, which is of practical importance for industrial users.

Numerical experiments on a mechanical system benchmark show that the local optimization enabled by SQP complements the Monte-Carlo sampling to detect rare worst-case configurations faster and more reliably. 
Then, by iteratively constructing a set of active configurations, multi-objectives robust controller optimization was performed in reasonable runtime.

\section*{Acknowledgment}
The authors thank Vincent Guibert for his valuable feedback on the preprint.

\section*{Declaration of Generative AI and AI-assisted technologies in the writing process}
During the preparation of this work the authors used ChatGPT (GPT-5) to assist in improving the wording and reformulation of selected sentences. After using this tool, the authors reviewed and edited the content as needed and take full responsibility for the content of the publication.

\section*{Funding Sources}

This work was funded by Région Occitanie in the frame of the France 2030 program.
Ce projet a été financé par la Région Occitanie dans le cadre de France 2030.
Grand number:  DOS0259269/00.

\appendix

\section{Mathematical proofs}

In this Appendix, we prove the propositions of Section~\ref{sec:optim} that result from mere adaptations of existing literature.

\vspace{2mm}

\noindent \textbf{Proposition~\ref{prop:upperC1constrained}} \textbf{: Descent directions and optimality of upper-$C^1$ functions -- Constrained case}


    \noindent \emph{a) Descent direction. }

\begin{proof}
The proof of a) is an adaptation of \cite[Theorem 3.1]{Han1997}, originally written for a differentiable $f$, to the upper-$C^1$ case.
Let us define: 
\vspace{-2mm}
\begin{align*}
I = \lbrace i: c_i(x) > 0 \rbrace
\\
\bar I = \lbrace i: c_i(x) = 0 \rbrace
\\
\hat I = \lbrace i: c_i(x) < 0 \rbrace
\end{align*}
As in \cite{Han1997} we have:
\[
\theta'_r(x,p) = f'(x,p) + r \sum_{i \in I} \nabla c_i(x)^Tp  + r \sum_{i \in \bar I} (\nabla c_i(x)^Tp)_+ \;.
\]
Since $f$ is upper-$C^1$, from \cite[Proposition 2]{Oliveira2023} we have $f'(x,p) = \min_{g \in \partial f(x)} g^T p \leq g_{x}^Tp$ and thus
\[
\theta'_r(x,p) \leq g_{x}^Tp + r \sum_{i \in I} \nabla c_i(x)^Tp  + r \sum_{i \in \bar I} (\nabla c_i(x)^Tp)_+ \;.
\]
In the original proof of~\cite[Theorem~3.1]{Han1997}, the above expression appears as an equality rather than an inequality, with $\nabla f(x)$ in place of $g_x$. This minor difference does not affect the remainder of the proof. Indeed, since ($p,u$) is a KKT pair of $Q(x,H,g_{x})$, we have:
\[
g_{x} + \sum_{i=1}^n u_i \nabla c_i(x) + \frac{1}{2} (H+H^T)p = 0
\]
where, again, the original proof involves $\nabla f(x)$ in place of $g_x$. This relation is then used to substitute $g_x$ (or $\nabla f(x)$ in the original proof) in the right-hand side of previous inequality, together with additional equalities and inequalities involving only the constraints $c_i$, which remain unchanged in our setting. This eventually yields
\begin{equation*}
\theta'r(x,p) \leq - \frac{1}{2} p^T(H+H^T)p + \sum_{i \in I} (u_i -r ) c_i(x) \
< 0 \;.
\end{equation*}

\end{proof}

\noindent \textbf{Lemma 2: Perturbation of a quadratic problem}


\begin{proof}
    This lemma is an adaptation of \cite{Daniel1973} where, instead of our assumption $(k'-k)^T(x'_0-x_0) \leq \epsilon ||x'_0-x_0||$, the stronger assumption $||k'-k||\leq \epsilon$ was considered. 
    Let us note $\lambda >0$ the smallest eigenvalue of $K$.
    The proof of \cite{Daniel1973} starts with \cite[Theorem 2.1]{Daniel1973} that considers the unconstrained case:
    \[  x_0 = \arg \min \left\lbrace Q(x) = \frac{1}{2} x^T K x - x^T k \right\rbrace\]
    \[x'_0 = \arg \min \left\lbrace Q'(x) = \frac{1}{2} x^T K' x - x^T k' \right\rbrace \]
    and demonstrates that
        \begin{equation*}
    (\lambda - \epsilon) ||x'_0-x_0||^2 \leq \epsilon ||x'_0-x_0||(1+||x_0||) \;.
    \end{equation*}
    This results follows from intermediate results that are unmodified in our case, and from the inequality
    \begin{equation*}
        (x'_0-x_0)^T (k'-k)  \leq \epsilon ||x'_0-x_0||
    \end{equation*}
    that follows from $(x'_0-x_0)^T (k'-k)   \leq  ||x'_0-x_0||  \cdot || k' - k||$ and from their assumption $||k'-k||\leq \epsilon$. Our assumption directly implies the same inequality, so the proof of~\cite[Theorem~2.1]{Daniel1973} carries over without further modification.
    
    Then, the rest of the proof provided in \cite{Daniel1973} still holds. Indeed, the propositions \cite[Propositions 3.1, 3.2]{Daniel1973} only tackle the constrained sets, and do not involve $k$ or $k'$. Moreover, in, \cite[Section 4]{Daniel1973}, the relationship between $k$ and $k'$ is not used, as it only works on $Q(x)$.
\end{proof}

\noindent \textbf{Proposition~\ref{th:convergence} : Convergence of SQP for upper-$C^1$ functions}

    



\begin{proof}
    We follow the proof of \cite[Theorem 3.2]{Han1997} and adapt it to the case where $f$ is not differentiable but only upper-$C^1$. For completeness, we reproduce the entire proof and indicate where our modified assumptions apply, while emphasizing that the proof itself is not our own.

    By assumption (ii), $(p_k,u_k)$ is a KKT pair of $Q(x_k, H_k, g_{x_k})$ with $||u_k||_\infty \leq r$. If $p_k=0$, then $(x_k,u_k)$ satisfies the KKT conditions~\eqref{eq:kktoptim} of Problem~\eqref{eq:upperC1problem} and the algorithm can terminate at $x_k$.

    Thus, for the following, we assume that $p_k \neq 0$. From Proposition~\ref{prop:upperC1constrained}.a) (originally \cite[Theorem 3.1]{Han1997}, which we adapted to upper-$C^1$ functions), it is possible to choose $x_{k+1}$ as described in Algorithm~\ref{sqp}, and in particular, it verifies:
    \[  
    \theta_r(x_{k+1}) < \theta_r(x_k) + \epsilon_k \;.
    \]
    Let $\bar x$ be an accumulation point of $(x_k)$ with $S^0(\bar x) \neq \emptyset$. Without loss of generality (passing to the subsequence if necessary) we may assume
    \[
    x_k \rightarrow \bar x \;, H_k \rightarrow \bar H
    \]
    where the existence of $\bar H$ follows from assumption (i). 
    
    In the original proof \cite{Han1997}, the continuity of the gradient of $f$ ensures that $\nabla f(x_k)$ converges to $\nabla f(\bar x)$. Here, a similar property is maintained as follows. Since $f$ is locally Lipschitz, its subgradients $(g_{x_k})$ are uniformly bounded for $k$ sufficiently large, hence there exists a subsequence that converges to a vector $\bar g$. Without loss of generality (passing to the subsequence if necessary) we note
    \[  g_{x_k} \rightarrow \bar g \;,
    \]
    and the upper semi-continuity of the Clarke subdifferential \cite[Proposition 1.5]{Clarke1998} yields $\bar g \in \partial f(\bar x)$.
    Since $\bar H$ is positive definite and $S^0(\bar x) \neq \emptyset$, the quadratic problem $Q(\bar x, \bar H, \bar g)$ (originally $Q(\bar x, \bar H, \nabla f(\bar x))$ in \cite{Han1997}) has a unique KKT point~$\bar p$. If $\bar p =0$, then $\bar x$ is a KKT point of Problem~\eqref{eq:upperC1problem} and the theorem follows. Thus, we now assume that $\bar p \neq 0$ and we will show that this leads to a contradiction.

    From the continuity of $\nabla c(x)$, and by Lemmas 1 and 2, where $x_{k+1}$ and $x_k$ play the roles of $x_0$ and $x'_0$, and $g_{x_0}$ and $g_{x'_0}$ play the roles of $k$ and $k'$, we deduce that:
    \[
    p_k \rightarrow \bar p
    \]
    In the original proof \cite{Han1997}, this instead follows from \cite{Daniel1973} (the equivalent to our Lemma~2) and from the continuity of $\nabla f(x)$ (which is stronger than our Lemma~1).

    Since $(u_k)$ is uniformly bounded by $r$, it has an accumulation point $\bar u$. We note
    \[
    u_k \rightarrow \bar u
    \]
    with $||\bar u||_\infty \leq r$. By passing to the limit in the KKT relation, which is valid for any $k$, we obtain that $\bar u$ is a Lagrange multiplier of $Q(\bar x, \bar H, \bar g)$.

    The remainder of the proof of~\cite{Han1997} is unchanged and is reproduced below for completeness. Let $\bar \lambda \in [
        0 , \;  \lambda_{max}]$ be chosen such that 
    \[
    \theta_r(\bar x + \bar \lambda \bar p) = \min_{0 \leq \lambda \leq \lambda_{max }} \; \theta_r(\bar x + \lambda \bar p). 
    \]
    Let us set
    \[
    \gamma = \theta_r(\bar x + \bar \lambda \bar p) - \theta_r(\bar x) > 0
    \]
    which is strictly positive by Proposition~\ref{prop:upperC1constrained}.a). Since $x_k + \bar \lambda p_k \rightarrow \bar x + \bar \lambda \bar p$, for sufficiently large $k$ we have:
    \[
    \theta_r(x_k + \bar \lambda p_k) + \gamma/2 < \theta_r(\bar x) \;.
    \]
    However, for all $k$ we have $\theta_r(x_{k+1})<\theta_r(x_k) + \epsilon_k$, and for sufficiently large $k$ we have
    \[
    \sum_{i=k}^\infty \epsilon_i < \gamma/2
    \]
    which leads to:
    \begin{equation*}
    \begin{aligned}
    \theta_r(\bar x) &< \theta_r(x_{k+1}) + \sum_{i=k+1}^\infty \epsilon_i \\
    & \leq \min_{0 \leq \lambda \leq \lambda_{max}} \theta_r(x_k + \lambda p_k) + \epsilon_k +  \sum_{i=k+1}^\infty \epsilon_i \\
    &<  \theta_r(x_k + \bar \lambda p_k) + \gamma/2
    \end{aligned}
    \end{equation*}
    which contradicts the previous inequality $\theta_r(x_k + \bar \lambda p_k) + \gamma/2 < \theta_r(\bar x)$.
    This proves that $\bar p=0$ and that $\bar x$ is a KKT point of Problem~\eqref{eq:upperC1problem}.
\end{proof}

\bibliographystyle{IEEEtran}
\bibliography{library}

\end{document}